\documentclass[conference]{IEEEtran}
\IEEEoverridecommandlockouts
\usepackage{cite}
\usepackage{amsmath,amssymb,amsfonts}
\usepackage{algorithm}
\usepackage{algpseudocode}
\usepackage{verbatim}
\usepackage{graphicx}
\usepackage{setspace}
\usepackage{textcomp}
\usepackage{xcolor}
\usepackage{caption}
\usepackage{subcaption}
\usepackage{varwidth}
\usepackage{tikz}
\usepackage{flushend}
\usetikzlibrary{tikzmark}
\usepackage{multirow}

\definecolor{mygreen}{RGB}{0,0,0}
\definecolor{myblue}{RGB}{0,0,0}
\definecolor{myred}{RGB}{0,0,0}
\definecolor{mymagenta}{RGB}{0,0,0}

\def\BibTeX{{\rm B\kern-.05em{\sc i\kern-.025em b}\kern-.08em
    T\kern-.1667em\lower.7ex\hbox{E}\kern-.125emX}}
\algrenewcommand\algorithmicrequire{\textbf{Input:}}
\algrenewcommand\algorithmicensure{\textbf{Output:}}

\newcommand{\abd}{\boldsymbol{a}}
\newcommand{\bbd}{\boldsymbol{b}}
\newcommand{\cbd}{\boldsymbol{c}}

\newcommand{\fbd}{\boldsymbol{f}}

\newcommand{\obd}{\boldsymbol{o}}
\newcommand{\pbd}{\boldsymbol{p}}

\newcommand{\sbd}{\boldsymbol{s}}

\newcommand{\wbd}{\boldsymbol{w}}
\newcommand{\xbd}{\boldsymbol{x}}
\newcommand{\ybd}{\boldsymbol{y}}
\newcommand{\zbd}{\boldsymbol{z}}

\newcommand{\Abd}{\boldsymbol{A}}
\newcommand{\Bbd}{\boldsymbol{B}}

\newcommand{\Lbd}{\boldsymbol{L}}

\newcommand{\Rbd}{\boldsymbol{R}}

\newcommand{\Ybd}{\boldsymbol{Y}}

\newcommand{\Sigmabd}{\boldsymbol{\Sigma}}

\DeclareMathOperator*{\argmax}{arg\,max}

\begin{document}

\title{Parallel Approximations for High-Dimensional Multivariate Normal Probability Computation in Confidence Region Detection Applications
}

\author{\IEEEauthorblockN{Xiran Zhang$^{1,2}$, Sameh Abdulah$^{1,3}$, Jian Cao$^{4}$, Hatem Ltaief$^{1,3}$, Ying Sun$^{1,2,3}$,\\ Marc G. Genton$^{1,2,3}$, and David E. Keyes$^{1,3}$}
\IEEEauthorblockA{$^1$\textit{Computer, Electrical, and Mathematical Sciences and Engineering Division, King Abdullah University of Science and Technology,}\\
$^2$\textit{Statistics Program, King Abdullah University of Science and Technology,}\\ 
$^3$\textit{Extreme Computing Research Center, King Abdullah University of Science and Technology,}\\
\textit{Thuwal 23955-6900, Saudi Arabia.} \\
$^4$\textit{Department of Mathematics, The University of Houston, Houston, Texas, USA.}}}


\maketitle

\begin{abstract}
Addressing the statistical challenge of computing the multivariate normal (MVN) probability in high dimensions holds significant potential for enhancing various applications. For example, the critical task of detecting confidence regions where a process probability surpasses a specific threshold is essential in diverse applications, such as pinpointing tumor locations in magnetic resonance imaging (MRI) scan images, determining hydraulic parameters in groundwater flow issues, and forecasting regional wind power to optimize wind turbine placement, among numerous others. One common way to compute high-dimensional MVN probabilities is the Separation-of-Variables (SOV) algorithm. This algorithm is known for its high computational complexity of $O(n^3)$ and space complexity of $O(n^2)$, mainly due to a Cholesky factorization operation for an $n \times n$ covariance matrix, where $n$ represents the dimensionality of the MVN problem. This work proposes a high-performance computing framework that allows scaling the SOV algorithm and, subsequently, the confidence region detection algorithm. The framework leverages parallel linear algebra algorithms with a task-based programming model to achieve performance scalability in computing process probabilities, especially on large-scale systems. In addition, we enhance our implementation by incorporating Tile Low-Rank (TLR) approximation techniques to reduce algorithmic complexity without compromising the necessary accuracy. To evaluate the performance and accuracy of our framework, we conduct assessments using simulated data and a wind speed dataset. Our proposed implementation effectively handles high-dimensional multivariate normal (MVN) probability computations on shared and distributed-memory systems using finite precision arithmetics and TLR approximation computation. Performance results show a significant speedup of up to 20X in solving the MVN problem using TLR approximation compared to the reference dense solution without sacrificing the application's accuracy. The qualitative results on synthetic and real datasets demonstrate how we maintain high accuracy in detecting confidence regions even when relying on TLR approximation to perform the underlying linear algebra operations.
\end{abstract}


\section{Introduction}
The multivariate normal distribution extends the concept of the univariate normal distribution to encompass higher dimensions. In the context of probability, it characterizes the distribution of a random vector featuring multiple components, each potentially correlated with the others. The applications of multivariate normal probability span diverse fields, such as its use in spatial statistics for climate modeling~\cite{cannon2018multivariate} and confidence region detection~\cite{bolin2015excursion}, its application in machine learning through Gaussian Mixture Models (GMMs)~\cite{pignat2019bayesian} and Principal Component Analysis (PCA)~\cite{sun2021machine}, and its role in economic studies, particularly in macroeconomic modeling~\cite{amisano2017prediction}.

Confidence region detection is a well-established problem primarily focused on identifying spatial areas where a given process exceeds a certain threshold with a given probability. For example, in applications related to air pollution, identifying areas where pollution levels exceed a specific value, posing a potential risk to human health, and ensuring accurate assessment and management of environmental risks~\cite{cameletti2013spatio}. Another example involves brain imaging applications that pinpoint specific brain regions exhibiting particular characteristics or responses~\cite{ejaz2022confidence}. The usage of confidence region detection can be extended to applications in spatial statistics, astrophysics~\cite{bevins2022astrophysical}, machine learning~\cite{ejaz2020hybrid}, environmental science~\cite{sommerfeld2018confidence}, and many others. Mathematically, given a latent stochastic field $X(\sbd)$ at spatial location $\sbd$ and a set of observations denoted as $\mathbf{y}$, the goal is to identify a region, denoted as $D$, where the condition $X(\sbd) > u$ holds true for $\sbd$ within $D$ with a probability of at least $1-\alpha$. Here, $u$ is the threshold value, and $1-\alpha$ is the confidence level. 

{\color{myblue}
A common strategy for addressing this statistical challenge involves computing the multivariate normal (MVN) probability,  which involves solving an integration problem in high dimensions. A Monte Carlo (MC) method can be used to solve this integration problem by simulating a large number of random samples and averaging the integral function over these samples~\cite{caflisch1998monte}. Nevertheless, using this MC method in high dimensions is practically prohibitive when accuracy is essential~\cite{bolin2015excursion}. In the literature, the Separation-of-Variables (SOV) algorithm has been used to transform the integration problem into a solvable format by transforming the integration region to a unit hypercube. Then, a quasi-MC method can be used to evaluate the new integration problem~\cite{cao2021exploiting,bolin2015excursion}.} The SOV algorithm is known for its computational complexity of $O(n^3)$ and space complexity of $O(n^2)$, with $n$ representing the dimensionality of the MVN problem. The high complexity arises due to the need for solving a Cholesky factorization problem for a given covariance matrix $\boldsymbol{\Sigma}$. With the substantial increase in the volume of spatial data originating from various sources, it becomes imperative to explore other approaches for solving the SOV problem in large dimensions. 

The availability of efficient algorithms for performing linear algebra operations on parallel architectures has enabled significant improvements in existing algorithms, making the handling of substantial scientific data across diverse domains feasible. Parallel tile-based linear solvers, finely tuned for modern hardware architectures, play a pivotal role in optimizing the execution of many applications, addressing previously unsolved problems. Notable parallel linear algebra library examples include Chameleon~\cite{chameleon-soft}, DPLASMA~\cite{bosilca2011flexible}, and HiCMA~\cite{akbudak2017tile,abdulah2019hierarchical}. The two formers support dense linear algebra matrix operations on shared and distributed-memory systems. The latter integrates support for Tile Low-Rank (TLR) matrix approximations to deal with large-scale scientific problems. Furthermore, leveraging dynamic runtime systems within these libraries contributes to the meticulous tuning of their execution on modern architectures, enabling higher rates of floating-point operations per second (flops) and efficient time-to-solution.

This study proposes a parallelized version of the SOV algorithm, effectively calculating the MVN probability in high-dimensional spaces. Our approach leverages Chameleon, which relies on the StarPU dynamic runtime system for efficiently managing large datasets using fine-grained computations on manycore systems. Additionally, we enhance our implementation to accommodate the TLR approximation for the underlying matrix operations via the HiCMA library. We show that our TLR implementation can reduce the complexity of the SOV algorithm and address challenges posed by high-dimensional MVN problems.

To assess the efficacy of our implementation, we focus on the confidence region detection applications that mainly rely on the MVN probability algorithm. Our results demonstrate that TLR efficiently replaces dense by TLR computations with acceptable accuracy loss. Through a comprehensive evaluation, we analyze the accuracy trade-offs associated with this approximation and compare them with dense computations using synthetic and real wind speed datasets.

\section{Contributions}

We position our paper against existing works \cite{bolin2015excursion, cao2021exploiting,cao2022tlrmvnmvt} and summarize our contributions as follows:
\begin{itemize}

    \item {\color{myblue} We leverage the sequential design for solving the Multivariate Normal (MVN) probability problem in high dimensions from~\cite{bolin2015excursion}. We develop an MVN parallel implementation targeting the SOV algorithm relying on the state-of-the-art parallel task-based linear algebra library Chameleon~\cite{chameleon-soft} powered by the StarPU dynamic runtime system~\cite{starpu}.}

    \item We enhance the SOV algorithm in two respects. Initially, we incorporate an optimized tile-based Cholesky factorization implementation into the algorithm, allowing faster execution on parallel architectures. Subsequently, we parallelize the computation of probabilities for different QMC samples by dividing the problem into independent tasks and implementing this process through the StarPU dynamic runtime system. {\color{myblue} Our proposed high-performance implementation contrasts with the {\em R} implementation in~\cite{cao2021exploiting,cao2022tlrmvnmvt} that has shown limitation in parallel efficiency due to the bulk synchronous programming model.}

    \item We extend our implementation to support parallel TLR approximation of the SOV algorithm, accommodating various compression accuracy levels to reduce the complexity of the SOV algorithm through the use of the {\color {myblue} HiCMA library~\cite{akbudak2017tile,abdulah2019hierarchical} and StarPU. Compared to~\cite{cao2021exploiting,cao2022tlrmvnmvt}, using the StarPU dynamic scheduler to orchestrate task parallelism enables us to mitigate the overhead of load imbalance.}

    \item We evaluate the performance and scalability of our proposed dense and TLR implementations on both shared- and distributed-memory systems. {\color{myblue} In this study, we process a problem size of approximately 500K and 760K in dense and TLR formats, respectively, which is a significant advancement compared to the roughly 16K by~\cite{cao2021exploiting}.}

    \item {\color {myblue} We improve the confidence region detection algorithm in~\cite{bolin2015excursion} by incorporating the high-performance MVN probability implementation. We further deploy it across diverse synthetic and real datasets.}

    \item We evaluate the reduction in accuracy when depending on TLR approximation in contrast to the dense solution, using synthetic datasets that depict varying levels of spatial correlation. 
    
    \item {\color{myblue} We use our new high-performance confidence region detection implementation} on a spatial wind speed dataset in the Middle East to identify optimal locations for establishing wind farms for energy production. The TLR-based algorithm demonstrates comparable accuracy in region detection compared to the dense execution.
\end{itemize}

\section{Background}
This section provides an overview of key terms used in this paper, including Multivariate Normal (MVN) probability, Separation-Of-Variables (SOV) algorithm, confidence region detection applications, task-based parallelism, parallel linear algebra libraries powered by
dynamic runtime systems, and Tile Low-Rank (TLR) approximation.

\subsection{Multivariate Normal (MVN) Probability}
A Multivariate Normal (MVN) probability is a statistical concept related to multivariate analysis, specifically dealing with the distribution of multiple variables. It is widely used in statistical applications, including Bayesian probit models~\cite{anceschi2023bayesian}, confidence region detection problems~\cite{bolin2015excursion}, and maximum likelihood estimation~\cite{cao2021exploiting}. It can be defined as a numerical integration problem with high computation complexity in higher dimensions. The MVN probability is defined as follows:

\vspace{-7mm}
\begin{multline}
     \Phi_n({\abd}, {\bbd}; {\boldsymbol \mu}, {\boldsymbol \Sigma}) = \\ 
     \int_{\abd}^{\bbd} \frac{1}{\sqrt{(2\pi)^n|{\boldsymbol \Sigma}|}} 
     \exp\Bigl\{ -\frac{1}{2} ({\xbd}-{\boldsymbol \mu})^\top {\boldsymbol \Sigma}^{-1} ({\xbd}-{\boldsymbol \mu})\Bigl\} \mbox{d}{\xbd}
     \label{eq:q1}
\end{multline}
where $\abd$ and $\bbd$ are two $n$-dimensional vectors representing the lower and upper integration limits, ${\boldsymbol \mu}$ representing the mean, and ${\boldsymbol \Sigma}$ the covariance matrix of a multivariate Gaussian distribution. Here ${\boldsymbol \Sigma}$ is constructed through a predetermined covariance function ${\boldsymbol \Sigma}_{ij}=C(\|{\bf h}_{ij}\|; {\boldsymbol \theta})$ where $\|{\bf h}_{ij}\|$ represents the distance between spatial locations $i$ and $j$, and ${\boldsymbol \theta}$ represents the statistical parameters of the underlying statistical field. Without loss of generality, we set ${\boldsymbol \mu} = \bf 0$ in this paper for simplicity. {\color{myblue} A direct approach to address the integration problem in equation~\ref{eq:q1} involves employing an MC method, which depends on random sampling to approximate and simulate the integration. This is achieved by generating a large number of random samples and then averaging the integration function values. However, if accuracy is a concern, this approach becomes impractical when dealing with high-dimensional problems, such as the one addressed in this study.}

In the literature, computing the MVN probability is a challenging task where quadrature-based algorithms are impractical in higher dimensions. Thus, 
Genz has proposed a Monte Carlo simulation solution to provide a numerical solution to the high-dimensional MVN probability~\cite{genz1992numerical}. The provided solution aims to transform the MVN problem into classic numerical integration problems that can be solved directly using standard integration algorithms. This transformation has been used in the literature under the name ``Separation-of-Variables (SOV)" algorithm. Several methods have been proposed in the literature based on the SOV algorithm for efficient calculation of the MVN probability; see~\cite{genz2009computation} for a review. One important research direction is to scale the  Monte Carlo solution to higher dimensions by using, for instance, the hierarchical decomposition of the covariance matrix to improve the computation time~\cite{genton2018hierarchical,cao2021exploiting}.

\subsection{Separation-Of-Variable (SOV) Algorithm}
In~\cite{genz1992numerical}, Genz has provided in detail the required transformation to equation~\ref{eq:q1} to be able to compute the MVN probability:

\vspace{-4mm}
\begin{multline}
\Phi_n(\abd, \bbd;{\bf 0}, {\bf \Sigma}) = 
\int_{\Phi(a_1^{\prime})}^{\Phi(b_1^{\prime})}
\int_{\Phi(a_2^{\prime})}^{\Phi(b_2^{\prime})}
\ldots
\int_{\Phi(a_n^{\prime})}^{\Phi(b_n^{\prime})}
\mbox{d}{\zbd},
     \label{eq:sov1}
\end{multline}
where $a_i^{\prime}=\frac{a_i-\Sigma_{j=1}^{i-1}\boldsymbol{L}_{ij}y_{j}}{\boldsymbol{L}_{ii}}$ and $b_i^{\prime}=\frac{b_i-\Sigma_{j=1}^{i-1}\boldsymbol{L}_{ij}y_{j}}{\boldsymbol{L}_{ii}}$. If we define $w_i \overset{i.i.d}{\sim} U(0,1)$ as random numbers from unit uniform distribution, then the transformation includes $\boldsymbol x={\boldsymbol {Ly}}$ and $y_i=\Phi^{-1}[\Phi(a_i^{\prime})+\{\Phi(b_i^{\prime})-\Phi({a_i^{\prime})}\}w_i]$.   Herein, ${\boldsymbol L}$ is the lower Cholesky factor of ${\boldsymbol \Sigma}$, i.e., ${\boldsymbol \Sigma}={\boldsymbol{ LL}}^\top$ and $\boldsymbol{L}_{ij}$ represents the $(i, j)$ element of $\boldsymbol{L}$. Computing the Cholesky factor $\boldsymbol{L}$ from $\boldsymbol{\Sigma}$ requires $O(n^3)$ operations and $O(n^2)$ memory, where $n$ represents the MVN dimension. The computation of the MVN integration with $\boldsymbol{w}$ is called Monte Carlo sampling~\cite{cao2022tlrmvnmvt}.  Detailed Monte Carlo algorithms description can be found in ~\cite{genz1992numerical,genz2009computation, cao2021exploiting}. We assume the field has a zero-mean function (i.e., ${\boldsymbol \mu} = \bf 0$).

To apply the SOV results to Monte Carlo algorithms, we have to transform equation \ref{eq:sov1} into
\begin{equation}
\begin{split}
\Phi_n(\abd, \bbd; \boldsymbol 0, \boldsymbol{\Sigma})
&= (b_1'-a_1') \int_{0}^{1} (b_2'-a_2')\\
&\cdots \int_{0}^{1} (b_n'-a_n') \int_{0}^{1} \mathrm d{\wbd}.
\end{split}
\label{eq:sov2}
\end{equation}
Then, we can obtain the probabilities in higher dimensions with random numbers in $[0,1]$, the univariate cumulative normal distribution function and its inverse.






\subsection {Confidence Region Detection}
In spatial statistics, confidence region (also known as excursion set) detection is the process of identifying areas in a given spatial region where the values exceed a certain level, i.e., threshold $u$, or expected range with a certain level of confidence $1-\alpha$. Assuming a random field $X$, we can define the confidence set as shown in~\cite{bolin2015excursion} by:

\begin{equation}
E_{u,\alpha}^+(X) = \argmax_D\left\{ |D|: \mathbb P\left\{ D \subseteq A_u^+(X) \right\} \ge 1 - \alpha \right\},
\label{eq:confidence_set}
\end{equation}
where $A_u^+(X) = \{\sbd \in \Omega: X(\sbd) > u\}$ and $\Omega$ is the spatial domain. The positive confidence function {\color{mymagenta}can be defined as:}
\begin{equation}
F_u^+(\sbd) = \sup \{1-\alpha; \sbd \in E_{u,\alpha}^+ (X)\}.
\label{eq:confidence_func}
\end{equation}
The confidence region can be easily computed using the confidence function as $\{\sbd: F_u^+(\sbd) \ge 1-\alpha\}$.

This process is used in many applications to locate places where data points exceed predefined limits or exhibit unusual patterns. Some examples are identifying the levels of air pollution in a specific region~\cite{cameletti2013spatio} and recognizing tumors in MRI images~\cite{ejaz2022confidence}.

The mathematical definition of the confidence region/excursion set detection problem is as follows: given a latent stochastic field $x(\mathbf{s})$ and a set of observations denoted as $\mathbf{y}$, our goal is to identify the region $E_{u,\alpha}^+$. Here, the threshold value $u$ and confidence level $1-\alpha$ are user-defined values.
This $E_{u,\alpha}^+$ can be approximately captured using marginal probabilities as shown in~\cite{bolin2015excursion}, which needs to be more precise to detect the correct regions. Hence, in the work by Bolin and Lindgren~\cite{bolin2015excursion}, an algorithm is proposed, focusing on the computation of Multivariate Normal (MVN) probability for detecting confidence regions in spatial data. However, the complexity of the MVN probability algorithm poses limitations, rendering it unsuitable for handling large spatial areas, a necessity in numerous applications.


\subsection {Task-based Linear Algebra Libraries and Dynamic Runtime Systems}
Task-based parallelism is a parallel computing paradigm that deals with the target problem as a collection of tasks with predetermined dependencies. These tasks can execute concurrently when computing resources are available and as long as there are no violations of the pre-established dependencies. Task-based parallelism offers flexibility and fine-grained computations, making it suitable for various parallel computing applications. Thus, cutting-edge parallel dense linear algebra libraries, like Chameleon~\cite{chameleon-soft} and DPLASMA~\cite{bosilca2011flexible}, utilize task-based parallelism to deliver efficient and dependable parallel linear solvers using meticulously designed tile-based algorithms. Additionally, harnessing dynamic runtime systems like StarPU~\cite{starpu}, PaRSEC~\cite{parsec-cise}, QUARK~\cite{yarkhan2011quark}, and others can enhance task management and scheduling on available resources, considering workload and resource availability for improved performance. In this work, we rely on the StarPU dynamic runtime system because of its high level of user-productivity achieved via abstraction for expressing parallelism, simplifying the development of parallel applications. It also includes scheduling heuristics for optimizing data movement between different memory hierarchies, which is crucial for performance optimization.

\subsection {Tile Low-Rank (TLR) Approximation}
The low-rank approximation is a prevalent mathematical technique used to estimate a dense matrix by representing it as the product of one or more matrices with lower ranks. Given the widespread use of tile-based algorithms in numerous linear algebra packages, Akbudak et al. have introduced a Tile Low-Rank (TLR) approximation method for manycore systems~\cite{akbudak2017tile}. This approach enables the separate approximation of each tile using the Singular Value Decomposition (SVD) algorithm. In this method, the ranks of the tiles correspond to the most significant singular values and vectors within each off-diagonal tile. The effectiveness of the TLR technique hinges on the ranks achieved for the off-diagonal tiles after compression, which, in turn, is influenced by the precision requirements of the specific application. The use of TLR approximation has found application in various domains, enabling efficient compression of dense matrices and rapid execution of underlying linear algebra operations with acceptable accuracy. Examples include climate modeling~\cite{akbudak2017tile,abdulah2018tile, mondal2023tile}, astronomy~\cite{ltaief2021meeting}, and seismic computation~\cite{ltaief2023scaling}.

\section{Confidence Region Detection Framework}
This section introduces the proposed computational framework for the confidence region detection problem. By considering a collection of spatial locations, their corresponding measurements, a user-defined threshold $u$, and a user-defined confidence level $1-\alpha$, the proposed framework can identify regions where values surpass the defined threshold $u$. We also demonstrate our contribution in parallelizing the underlying high-dimensional Multivariate Normal (MVN) probability algorithm. For clarity, we summarize all symbols used in this section in Table~\ref{tab:symbols} to facilitate understanding of the content. We use regular (unbolded) characters to denote scalar values, bold lowercase characters for vectors, and bold uppercase characters for matrices. We also use $\boldsymbol{A}(i,j)$ to denote the $(i,j)$ element in a matrix $\boldsymbol{A}$, and $\boldsymbol{A}_{(i,j)}$ to indicate the $(i,j)$ tile in the matrix $\boldsymbol{A}$.
\begin{table}[h]
\caption{List of symbols.}
\centering
\begin{tabular}{cl}
\hline
Symbol & Definition \\
\hline
$\boldsymbol{geom}$ & A set of irregularly distributed spatial locations.\\
$\boldsymbol{\hat\theta}$ & Estimated statistical parameter vector.\\
$1-\alpha$ & User-defined confidence level. \\
$u$ & User-defined threshold. \\
$\boldsymbol{\Sigma}$ & Covariance matrix.\\
$\Phi(x)$ & Univariate normal distribution function $\mathbb P(Z \le x)$.\\
$\pbd_{M}$ & Marginal probability vector $\mathbb P(X_i>u)$. \\
$\obd_{\pbd_{M}}$ & Indices of ordered $\pbd_{M}$.  \\
$\fbd$ & Positive confidence function in equation \ref{eq:confidence_func}. \\
$\abd$ & Lower limits of the MVN integration. \\
$\bbd$ & Upper limits of the MVN integration. \\
$\Abd$ & Matrix of $n$ set of $\abd$ vectors. \\
$\Bbd$ & Matrix of $n$ set of $\bbd$ vectors. \\
$\Rbd$ & Random matrix filled with i.i.d $U(0,1)$ values. \\
\hline
\end{tabular}
\label{tab:symbols}
\end{table}

\subsection{Confidence Region Detection Algorithm}
Algorithm~\ref{alg:excusion_set} illustrates how to identify confidence regions relying on the computation of Multivariate Normal (MVN) probabilities. The algorithm takes several inputs, as indicated in line 1, including $\boldsymbol{\hat\theta}$ obtained using a specific covariance function of the form $C(\|\mathbf{h}\|; \boldsymbol{\theta})$, where $\mathbf{h} = \mathbf{s}_1 - \mathbf{s}_2 \in \mathbb{R}^d$ and $\| \mathbf{h} \|$ denotes the Euclidean norm. Herein, we employ the Mat\'ern covariance function~\cite{gneiting2010matern} with the form:  
\begin{equation}\label{eqn:multivariate_matern}
C(\|\mathbf{h}\|; \boldsymbol{\theta}) = \frac{\sigma^2}{2^{\nu - 1} \Gamma \left( \nu \right)} \left(\frac{\|\mathbf{h}\|}{a}\right)^{\nu} \mathcal{K}_{\nu}\left(\frac{\|\mathbf{h}\|}{a}\right),
\end{equation}
where $\mathcal{K}_{\nu}(\cdot)$ represents the modified Bessel function of the second kind with order $\nu$, and $\Gamma(\cdot)$ denotes the gamma function. In this context, $\boldsymbol{\theta}$ contains the marginal variance ($\sigma^2>0$), smoothness ($\nu>0$), and spatial range ($a>0$). We obtained the Mat\'ern covariance parameters $\boldsymbol{\hat\theta}$ for the given dataset using the Maximum Likelihood Estimation (MLE) algorithm employed in the {\it ExaGeoStat} software~\cite{abdulah2018exageostat}.

\begin{algorithm}
\caption{Confidence Region Detection Algorithm}
\begin{algorithmic}[1]
\Function{\lowercase{CRD} }{$\boldsymbol{geom}$: a set of spatial locations, $\boldsymbol{Y}$ a set of measurements of sampled locations, $\boldsymbol{\hat\theta}$: Mat\'ern covariance parameters; 
$n$: number of spatial locations}
    \State Generate a covariance matrix $\boldsymbol{\Sigma}$ using $\boldsymbol{\hat\theta}$ and $\boldsymbol{geom}$ or read a given covariance matrix $\Sigmabd$.
    \For{$0 \le i < n$}
        \State $\pbd_M[i] \gets 1 - \Phi((u-\mu[i]-\Ybd[i])/\sqrt{\boldsymbol{\Sigma}[i,i]})$
    \EndFor
    \State $\obd_{\pbd_M} \gets \mathrm{descending\_order\_index}(\pbd_{M})$
    \State $\mathrm{pmvn\_init}()$
    \State \tikz[remember picture] \node[inner sep=0pt] (a) {$\Lbd$ $\gets$ dpotrf($\boldsymbol{\Sigma}$) \hspace{17mm}\Comment{Cholesky factorization}};
    \State $\bbd \gets \{\infty,\dots,\infty\}$
    \For{$n > i \ge 0$}
        \State $\cbd \gets \obd_{\pbd_M}[1:i]$
        \State $\abd \gets \{-\infty,\dots,-\infty\}$
        \State $\abd[\cbd] \gets (u-\mu[\cbd]-\Ybd[\cbd])/\sqrt{\boldsymbol{\Sigma}[\cbd,\cbd]}$
        
        \State $\fbd[\cbd] \gets \mathrm{PMVN}(\abd, \bbd, \boldsymbol{L}, n, N, m)$
    \EndFor
\EndFunction

\end{algorithmic}
\label{alg:excusion_set}
\end{algorithm}

In line 2, the algorithm generates a covariance matrix $\boldsymbol{\Sigma}$ using the estimated parameters $\boldsymbol{\hat\theta}$ and the set of locations $\boldsymbol{geom}$. In lines 3-5, the marginal probabilities of the locations denoted as $\pbd_M$, are calculated using the mean $\mu_i$, i.e., $\mu_i$ is the mean in location $i$ across time, where $u$ is a user-defined threshold value. In line 6, we store the indices of the locations based on their marginal probabilities in $\obd_{\pbd_M}$ vector in descending order. In line 7, the algorithm initializes the required data structures that will be used in the computation, i.e., Chameleon/HiCMA descriptors. The descriptors are unique data structures that enable the storage of matrices as a set of tiles managed by one or more computing processes. In the case of HiCMA, the pmvn\_init() function also encompasses the compression of the covariance matrix $\boldsymbol{\Sigma}$ into the TLR format. In line 8, a Cholesky factorization operation is performed over $\boldsymbol{\Sigma}$ to obtain the lower triangular matrix $\boldsymbol{L}$, where $\boldsymbol\Sigma=\boldsymbol{L}\boldsymbol{L}^\top$. In lines 9-15, the locations are extracted according to $\obd_{\pbd_M}$ and the joint MVN probabilities are computed using the PMVN function in algorithm~\ref{alg:pmvn} based on the lower and upper limits, constructing the confidence function $\fbd$.

\tikz[overlay,remember picture] {
  \draw[red,thick] ([shift={(-0.05,-0.05)}]a.south west) rectangle node[shift={(-0.7,0.0)}] {(a)} ([shift={(0.05,0.05)}]a.north east);
}

\subsection{Multivariate Normal Probability (PMVN) Algorithm}

Algorithm~\ref{alg:pmvn} calculates the Multivariate Normal (MVN) probability. To allow parallel implementation of the algorithm, four matrices -- $\boldsymbol{A}$, $\boldsymbol{B}$,  $\boldsymbol{R}$, and a temporary matrix $\boldsymbol{Y}$ -- are used.  Matrices $\boldsymbol{A}$ and $\boldsymbol{B}$ incorporate redundant lower limits vector $\boldsymbol{a}$ and upper limits vector $\boldsymbol{b}$, respectively (lines 2-3). Matrix $\boldsymbol{R}$ is set with values drawn from a unit uniform distribution (line 4). In lines 5-7, the algorithm invokes the $\mathrm{QMC}()$ (introduced later in Algorithm~\ref{alg:qmc}) to update the set of tiles in the first row of matrices $\boldsymbol{A}$, $\boldsymbol{B}$, and $\boldsymbol{Y}$. This loop can be executed concurrently, with each task handling a single tile. To propagate these changes to all the tiles in the next row, a set of GEMMs operations are concurrently applied, as shown in lines 10-13. The $\mathrm{QMC}()$ algorithm is once again invoked for all the tiles in this row, as demonstrated in lines 15-17. These procedures are repeated until the final row is reached, resulting in the $\boldsymbol{p}$ probability vector for each column of tiles. The final MVN probability, $p$, is computed as the mean of the $\boldsymbol{p}$ vector (line 19).

\begin{algorithm}
\caption{Multivariate Normal Probability (PMVN) Integration Algorithm}
\begin{algorithmic}[1]
\Function{PMVN}{$\abd$: lower limits; $\bbd$: upper limits; $\boldsymbol{L}$: Cholesky factor; $n$: dimension; $N$: QMC sample size; $m$: tile size}
\State $\Abd \gets [\abd,\abd,\dots,\abd]\in\mathbb R^{n\times N}$ 
\State $\Bbd \gets [\bbd,\bbd,\dots,\bbd]\in\mathbb R^{n\times N}$ 
\State $\Rbd\in\mathbb R^{n\times N}; \Rbd(i,j) \overset{i.i.d}{\sim} \mathcal U(0,1)$
\tikzmark{start1}
\For{$0 \le k < N/m$} \Comment{$\left \lceil N/m\right \rceil $ is \# tile-by-column}
   \State $\mathrm{QMC}(\Lbd_{(0,0)}, \Rbd_{(0,k)}, \Abd_{(0,k)}, \Bbd_{(0,k)}, \pbd_{(k)}, \Ybd_{(0,k)})$
\EndFor
\tikzmark{end1}
\For{$1 \le r < n/m$} \Comment{$\left \lceil n/m\right \rceil $ is \# tile-by-row}
    \For{$r \le j < n/m$}
    \tikzmark{start2}
     \For{$0 \le k < N/m$}
            \State $\boldsymbol{A}_{(j,k)} \gets$
            $\boldsymbol{A}_{(j,k)}-\boldsymbol{L}_{(j,r-1)} \cdot \boldsymbol{Y}_{(r-1,k)}$ 
            \State $\boldsymbol{B}_{(j,k)} \gets$
            $\boldsymbol{B}_{(j,k)}-\boldsymbol{L}_{(j,r-1)} \cdot \boldsymbol{Y}_{(r-1,k)}$ 
     \EndFor
     \tikzmark{end2}
    \EndFor
    \tikzmark{start3}
    \For{$0 \le k < N/m$}
        \State \begin{varwidth}[t]{\linewidth}
      $\mathrm{QMC}(\Lbd_{(r,r)}, \Rbd_{(r,k)}, \Abd_{(r,k)},$\par
        \hskip\algorithmicindent $\Bbd_{(r,k)}, \pbd_{(k)}, \Ybd_{(r,k)})$
      \end{varwidth}
    \EndFor
    \tikzmark{end3}
\EndFor
\State\Return $\mathrm{mean}(\pbd)$ 
\EndFunction
\end{algorithmic}
\label{alg:pmvn}
\end{algorithm}

\tikz[overlay,remember picture] {
  \draw[red,thick] ([shift={(-135pt,-3pt)}]pic cs:start1) rectangle node[shift={(0.0,-0.4)}] {(b)} ([shift={(190pt,-3pt)}]pic cs:end1);
}
\tikz[overlay,remember picture] {
  \draw[red,thick] ([shift={(-78pt,-3pt)}]pic cs:start2) rectangle node[shift={(0.0,-0.6)}] {(c)} ([shift={(144pt,-3pt)}]pic cs:end2);
}
\tikz[overlay,remember picture] {
  \draw[red,thick] ([shift={(-37pt,-3pt)}]pic cs:start3) rectangle node[shift={(0.0,-0.6)}] {(d)} ([shift={(100pt,-3pt)}]pic cs:end3);
}

\subsection{Quasi-Monte Carlo (QMC) Algorithm}
Equation \ref{eq:sov2} shows that the parallelization of the algorithm is feasible only across Monte Carlo (MC) chains. This limitation arises because $a_i'$, $b_i'$, and $y_i$ are interdependent in an iterative manner. Assuming tile-based $\Abd$, $\Bbd$, and $\Ybd$ matrices within a specific tile, each row relies on the preceding row for the update process.

Algorithm \ref{alg:qmc} provides a mechanism to update a single tile of $\pbd$ and $\Ybd$ while operating on $m$ MC chains in total. For simplicity, we assume all tiles have the same dimensions. The algorithm requires one tile each for the lower Cholesky factor $\Lbd$, the random matrix $\Rbd$, the lower and upper limits matrices $\Abd$ and $\Bbd$, and the probability vector $\pbd$, the random matrix $\Ybd$. 
In lines 3-7, {\color {mymagenta} we initialize the first row} in a given tile to compute $a_1'$, $b_1'$, and $y_1$ in equation \ref{eq:sov2}. In lines 8-15, we simultaneously complete $m$ steps for all the $m$ MC chains. The function does not provide a direct output, as the updates to $\pbd$ and $\Ybd$ occur within the process and are propagated to subsequent tiles during the subsequent steps in Algorithm~\ref{alg:pmvn}.

In Algorithms~\ref{alg:excusion_set} and~\ref{alg:pmvn}, red boxes were added over specific lines to highlight the integration of task-based parallel computation as follows: (a) Cholesky factorization, (b) and (d) parallel QMC computations, and (c) parallel GEMMs. The step (a) can be performed using both dense and TLR computations. However, (b), (c), and (d) are only performed in dense since $\boldsymbol{A}$ and $\boldsymbol{B}$ are non-admissable matrices. 

\begin{algorithm}[h]
\caption{QMC for MVN probabilities Algorithm}
\begin{algorithmic}[1]
\Function{QMC}{$\Lbd$: Cholesky factor tile; $\Rbd$: random matrix tile; $ \Abd$: lower limits tile; $\Bbd$: upper limits tile; $\pbd$: probability vector tile; $\Ybd$: temp matrix tile}
\State $m \gets \mathrm{nrow}(\Abd)$ \Comment{Tile size}
\For{$1 \le j < m$}
    \State ${\Abd}(0,j) \gets \frac{{\Abd}(0,j)}{{\Lbd}(0,0)}$; ${\Bbd}(0,j) \gets \frac{{\Bbd}(0,j)}{{\Lbd}(0,0)}$
    \State \begin{varwidth}[t]{\linewidth}
      $\Ybd(0,j) \gets \Phi^{-1} [\Rbd(0,j) \cdot$\par
        \hskip\algorithmicindent $\left\{\Phi({\Bbd}(0,j)) - \Phi({\Abd}(0,j))\right\}]$
      \end{varwidth}
    \State $ \pbd_j \gets \pbd_j \cdot \left\{\Phi({\Bbd}(0,j)) - \Phi({\Abd}(0,j))\right\}$
\EndFor
\For{$1 \le i < m$} 
    \For{$1 \le j < m$}
        \State $s \gets \Lbd(i, 1:(i-1)) \Ybd(1:(i-1),j)$ 
        \State $a' \gets \frac{{\Abd}(i,j) - s}{{\Lbd}(i,i)}$; $b' \gets \frac{{\Bbd}(i,j) - s}{{\Lbd}(i,i)}$
        \State ${\Ybd}(i,j) \gets \Phi^{-1} \left[\Rbd(i,j) \cdot \left\{\Phi(b') - \Phi(a')\right\}\right]$
        \State $ \pbd_j \gets \pbd_j \cdot \left\{\Phi(b') - \Phi(a')\right\}$
    \EndFor
\EndFor
\EndFunction
\end{algorithmic}
\label{alg:qmc}
\end{algorithm}


\section{Results}
This section evaluates the proposed implementation for the MVN probability algorithm under various objectives. Initially, we intend to gauge the performance in terms of time-to-solution of the dense and TLR implementations on shared- and distributed-memory systems. Subsequently, our focus is on assessing the accuracy of the proposed framework in the context of confidence region detection applications, encompassing both synthetic and real datasets. Lastly, our goal is to evaluate the accuracy of TLR approximation in conjunction with confidence region detection applications, considering the compression level necessary to maintain the required precision for the application.

\subsection{Environment Settings}
\label{sec:env_set}
We assess the performance of our proposed framework across various shared-memory architectures: a dual-socket 28-core Intel Icelake 6330 running at $2.00$ GHz, a dual-socket 64-core Intel Cascade Lake running at $2.30$ GHz, a dual-socket 128-core AMD Milan running at $2.00$ GHz, and a dual-socket 64-core AMD Naples running at $2.20$ GHz. In the distributed-memory experiments, we rely on a Cray XC40 system, Shaheen-II, at KAUST, with $6{,}174$ dual-socket 16-core Intel Haswell processors operating at $2.3$ GHz. Each node in this system is equipped with $128$ GB of DDR4 memory. The Shaheen system, boasting a total of $197{,}568$ processor cores, is backed by an aggregate memory of $790$ TB.

We compile our software using gcc v10.2.0 and link it with various libraries, including Chameleon, HiCMA, HWLOC v2.8.0, StarPU v1.3.9, Intel MKL v2020.0.166, and NLopt v2.7.0 for optimization.

\subsection{Datasets}

{\color{mygreen} We use {\it ExaGeoStat}~\cite{abdulah2018exageostat} to generate three spatial synthetic datasets from the exponential kernel with the range parameter equal to $0.033$ (weak correlation), $0.1$ (medium correlation) and $0.234$ (strong correlation). Each dataset contains $40{,}000$ data points.}
The generated data comprise a set of locations and corresponding measurements recorded at those locations. We follow the synthetic data generation process in~\cite{cao2022tlrmvnmvt}, where $6{,}250$ samples under the additive noise $\mathcal N(0,0.5^2)$ denoted as $\ybd$, are randomly selected from the original data {\color{mymagenta} for} the sake of computing the posterior covariance matrix $\Sigmabd_{\mathrm{post}}$ as follows:
\begin{equation}\label{eqn:q_post}
\Sigmabd_{\mathrm{post}}=(\Sigmabd^{-1}+(1/{0.5^2})\boldsymbol{A}^\top\boldsymbol{A})^{-1}
\end{equation}
where  ${\Abd} \in \mathbb R^{6250\times 40000}$ denotes the indicator matrix, $\Sigmabd$ is the covariance matrix of $\xbd$, and $\boldsymbol{\mu}_{\mathrm{post}}$ is the posterior mean of $\xbd$ which can be computed as:
\begin{equation}\label{eqn:mu_post}
\boldsymbol{\mu}_{\mathrm{post}}= \boldsymbol{\mu}+(1/{0.5^2})\Sigmabd_{\mathrm{post}} \boldsymbol{A}^\top(\ybd-\boldsymbol{A}\boldsymbol{\mu})
\end{equation}
where $\boldsymbol{\mu}$ denotes the mean of $\xbd$. The posterior covariance matrix $\Sigmabd_{\mathrm{post}}$ can be used as input to Algorithm~\ref{alg:excusion_set} (line 2) and the posterior mean $\boldsymbol{\mu}_{\mathrm{post}}$ can be used to update the $\abd$ vector in line 12.


We also present an analysis of wind speed data in Saudi Arabia from 2013 to 2016 by \cite{GIANIwind-dataset2020115085}. The dataset consists of hourly measurements aggregated into daily values across $53{,}362$ locations. The focus of the study is to provide a quantitative measure of wind speed on a specific day, specifically July 15, 2015. We performed postprocessing steps before applying the confidence region detection algorithm to the data. First, we computed the mean and standard deviation of the daily wind speed, followed by standardizing the average wind speed on the chosen day. We can then assume the wind field adheres to a stationary Gaussian random field characterized by a zero mean. Second, a Mat\'ern kernel is fitted using {\it ExaGeoStat} on the transformed field. The data and the estimated parameters are inputs to the confidence region detection Algorithm~\ref{alg:excusion_set} to delineate regions on the map with a $0.95$ probability of experiencing high wind speeds. The results include plots illustrating the original plot, excursion sets derived through the proposed methods, and the marginal probability map. The comparison reveals that the marginal probability map differs significantly from the collective excursion set, highlighting the importance of modeling using multivariate normal probabilities. Additionally, the excursion maps generated by the dense and TLR methods exhibit consistency, with TLR being favored due to its faster computation.

\begin{figure*}[h!]
     \centering
     \begin{subfigure}[b]{\textwidth}
         \centering
         \includegraphics[width=0.20\textwidth,page=1]{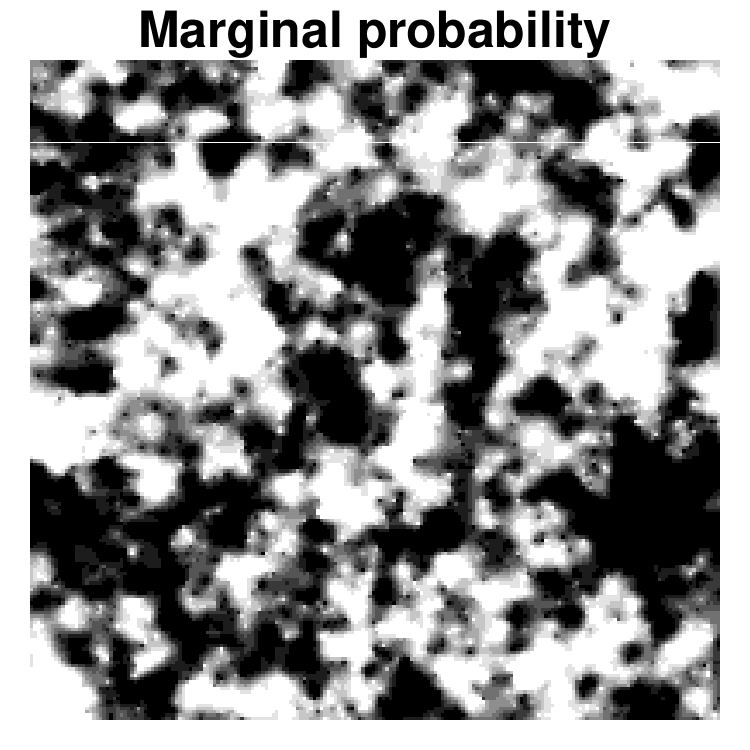}
            \hspace{4mm}
         \includegraphics[width=0.20\textwidth,page=2]{figures/Weak_correlation_plot.pdf}
          \hspace{4mm}
         \includegraphics[width=0.21\textwidth,page=3]{figures/Weak_correlation_plot.pdf}
                \hspace{4mm}
         \includegraphics[width=0.21\textwidth,page=4]{figures/Weak_correlation_plot.pdf}
         \caption{Weak correlation (1, 0.033, 0.5).}
         \label{fig:three sin x}
     \end{subfigure}
     \begin{subfigure}[b]{\textwidth}
         \centering
                \includegraphics[width=0.20\textwidth,page=1]{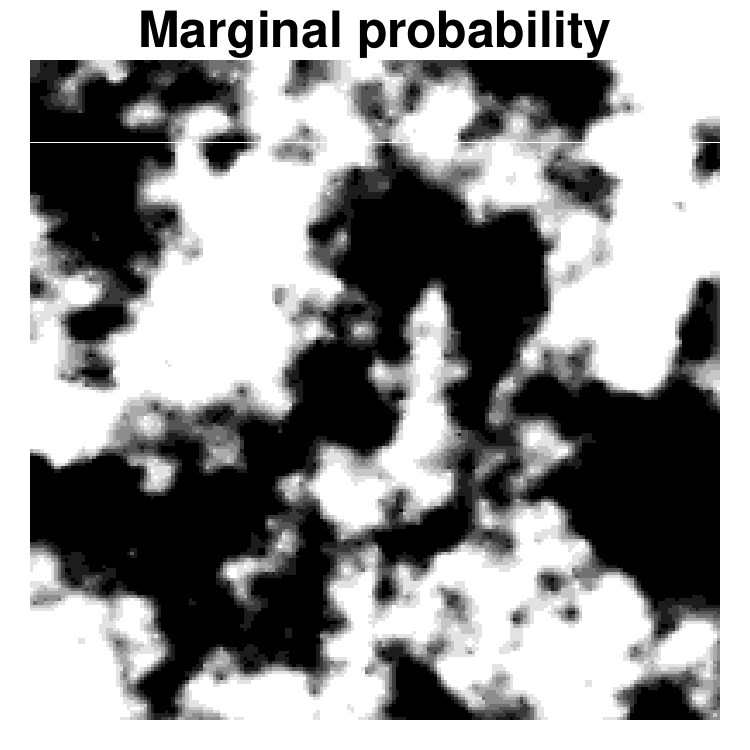}
            \hspace{4mm}
         \includegraphics[width=0.20\textwidth,page=2]{figures/Medium_correlation_plot.pdf}
          \hspace{4mm}
         \includegraphics[width=0.21\textwidth,page=3]{figures/Medium_correlation_plot.pdf}
                \hspace{4mm}
         \includegraphics[width=0.21\textwidth,page=4]{figures/Medium_correlation_plot.pdf}
         \caption{Medium correlation (1, 0.1, 0.5).}
         \label{fig:three sin x}
     \end{subfigure}
       \hspace{6mm}
     \begin{subfigure}[b]{\textwidth}
         \centering
                \includegraphics[width=0.20\textwidth,page=1]{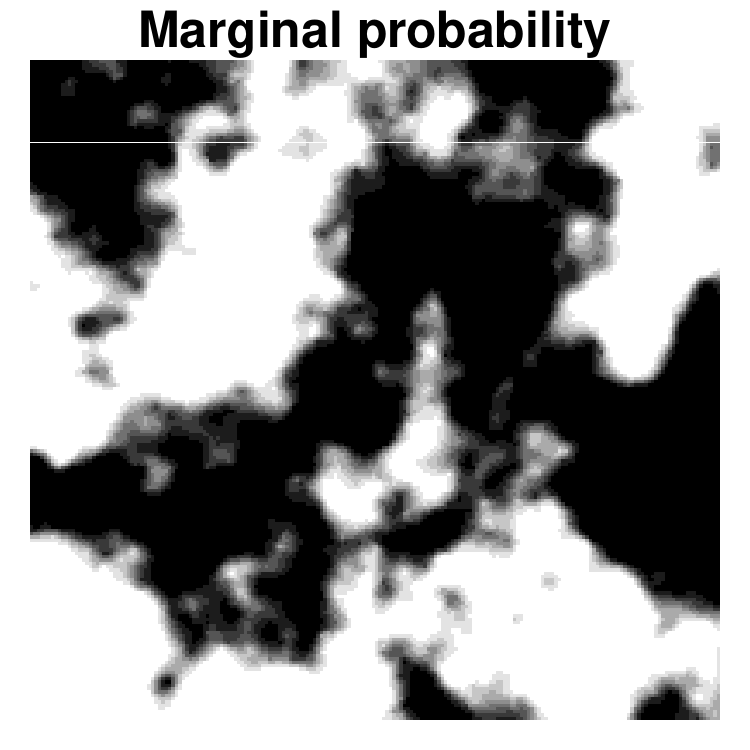}
            \hspace{4mm}
         \includegraphics[width=0.20\textwidth,page=2]{figures/Strong_correlation_plot.pdf}
          \hspace{4mm}
         \includegraphics[width=0.21\textwidth,page=3]{figures/Strong_correlation_plot.pdf}
                \hspace{4mm}
         \includegraphics[width=0.21\textwidth,page=4]{figures/Strong_correlation_plot.pdf}
         \caption{Strong correlation (1, 0.234, 0.5).}
         \label{fig:three sin x}
     \end{subfigure}
    \caption{Confidence region detection accuracy assessment using $40$K synthetic datasets generated in a regular grid with varying correlation levels. The figure illustrates the accuracy of both dense and TLR results compared to MC results (MC error) and the accuracy of TLR results compared to dense results.}
    \label{fig:acc_syn}
\end{figure*}

\subsection{Qualitative Assessment}
\label{sec:assess_acc}
We aim to evaluate the accuracy of identifying confidence regions using our proposed implementation for dense and TLR computations. We employed synthetic and real datasets, as elaborated in the following subsections. {\color {mygreen} 
All the experiments in this section rely on QMC sample size = $10{,}000$, which consistently yielded higher accuracy than smaller ones. In the performance section, we show the performance of our proposed implementations with different QMC sizes regardless of the obtained accuracy.}

\subsubsection{Accuracy Assessment on Synthetic Datasets}
For synthetic dataset accuracy assessment, we use three datasets with different properties, each consisting of $40{,}000$ data points, characterized by varying correlations, i.e., weak, medium, and strong. The results are depicted in Figure~\ref{fig:acc_syn}. For each correlation level, four images illustrate the detected regions using marginal probability, detected regions employing the confidence region detection algorithm, comparing dense and TLR results with MC results (MC error), and comparing TLR results with the dense results. {\color{myblue}
An MC validation algorithm can be used to validate the accuracy of confidence region detection methods, such as the ones proposed in this paper. This algorithm draws $N$ samples from the fitted distribution, where $N_s$ represents the number of samples exceeding a given threshold. The MC estimate of the confidence probability is then expressed as $\hat p(\alpha) = N_s/N$, and it is expected that $\hat p(\alpha) \approx 1 - \alpha$ if the confidence region is accurately estimated. In the third column of Figure 1, we show the difference $1-\alpha-\hat p(\alpha)$ as a function of $1-\alpha$, based on a sample size of $N=50{,}000$. The small discrepancy observed across all $\alpha$ values is primarily attributed to the error associated with the MC estimation of $\hat p(\alpha)$; this error is unrelated to the accuracy of our method.} The two left-row graphs reveal notable differences between the marginal probabilities and the confidence region obtained through the MVN algorithm. This underscores the significance of collectively evaluating the multivariate normal (MVN) probabilities. In particular, the confidence regions represent a subset where the joint probabilities exceed $1-\alpha$.

In the two right-row graphs, a comparison is made between probability results obtained from dense (red curves) and TLR (blue curves) methods with results derived from {\color{mymagenta}naive} MC chains. These results maintain the same level of error as those presented in \cite{cao2022tlrmvnmvt} and are also an order of magnitude lower than the second example in \cite{bolin2015excursion}, which utilizes an approximate posterior covariance matrix.

Additionally, we note that the difference between the results obtained from dense and TLR methods is minimal. To showcase these distinctions, we carried out experiments using TLR with varying levels of accuracy. For weak and medium correlations, an accuracy of $1e-1$ is acceptable, with discrepancies smaller than $1\times10^{-3}$. As accuracy increases, the gap between the two methods steadily diminishes in all instances. When accuracy surpasses $1e-3$, the difference becomes negligible enough to be ignored.

\subsubsection{Accuracy Assessment on a Real Wind Speed Dataset}

We performed some data preprocessing on the wind data before applying our algorithm, wherein we calculated the average daily wind speed for all summer days between 2013 and 2016 in Saudi Arabia. Next, we compute the mean and standard deviation of the summer wind speed over time. The July 15, 2015, wind speed record is then standardized by subtracting the mean and dividing by the standard deviation. The standardized data is subsequently fed into {\it ExaGeoStat} to estimate the Mat\'ern parameters, yielding the values $(1, 0.005069, 1.43391)$.

\begin{figure*}[h]
    \centering
    \begin{subfigure}[b]{0.245\textwidth}
         \centering
         \includegraphics[width=\textwidth,page=1]{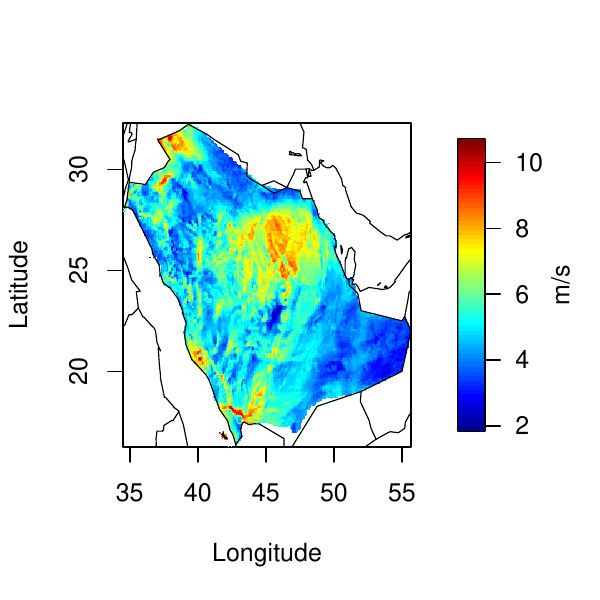}
         \caption{Original data.}
         \label{fig:real-data-set-a}
     \end{subfigure}
     \hfill
    \begin{subfigure}[b]{0.245\textwidth}
         \centering
         \includegraphics[width=\textwidth,page=1]{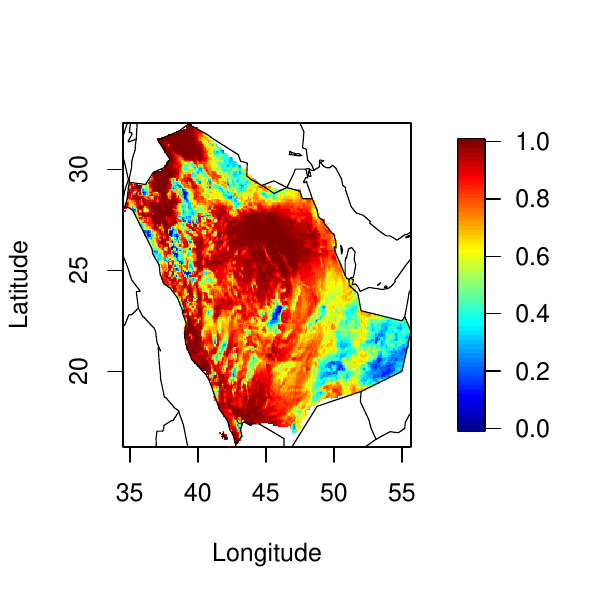}
         \caption{Marginal probability.}
         \label{fig:real-data-set-b}
     \end{subfigure}
     \hfill
    \begin{subfigure}[b]{0.245\textwidth}
         \centering
         \includegraphics[width=\textwidth,page=2]{figures/ws_plot_dense.pdf}
         \caption{Confidence regions (Dense).}
         \label{fig:real-data-set-c}
     \end{subfigure}
     \hfill
    \begin{subfigure}[b]{0.245\textwidth}
         \centering
         \includegraphics[width=\textwidth,page=2]{figures/ws_plot_TLR.pdf}
         \caption{Confidence regions (TLR).}
         \label{fig:real-data-set-d}
     \end{subfigure}
    \caption{Results of summer wind speed data (on July 15, 2015) in the Middle East (Saudi Arabia).}
    \label{fig:real-data-set}
\end{figure*}

In our analysis, we set the threshold value at $4$ m/s following \cite{ChenWanFangThreshold}. Furthermore, we opt for a confidence level of $0.95$. We employ a dense tile size of $320$, a TLR tile size of $980$, and set a maximum rank of $145$, maintaining the TLR accuracy level of $1e-4$. The experiment results are depicted in Figure~\ref{fig:real-data-set}. Figure~\ref{fig:real-data-set-a} illustrates the initial distribution of wind speeds on July 15, 2015, highlighting elevated wind speeds in the northern, eastern, and southwestern regions, varying from $2$ m/s to approximately $12$ m/s. Figure~\ref{fig:real-data-set-b} displays the marginal probability of wind speeds at various locations. However, the results pose a problem, as a significant portion of Saudi Arabia exhibits a probability greater than 0.8 of experiencing an average wind speed exceeding $4$ m/s on that particular day, which is highly unrealistic.

\begin{figure}[hbt!]
    \begin{subfigure}[b]{0.37\textwidth}
         \centering
         \includegraphics[width=\textwidth,page=1]{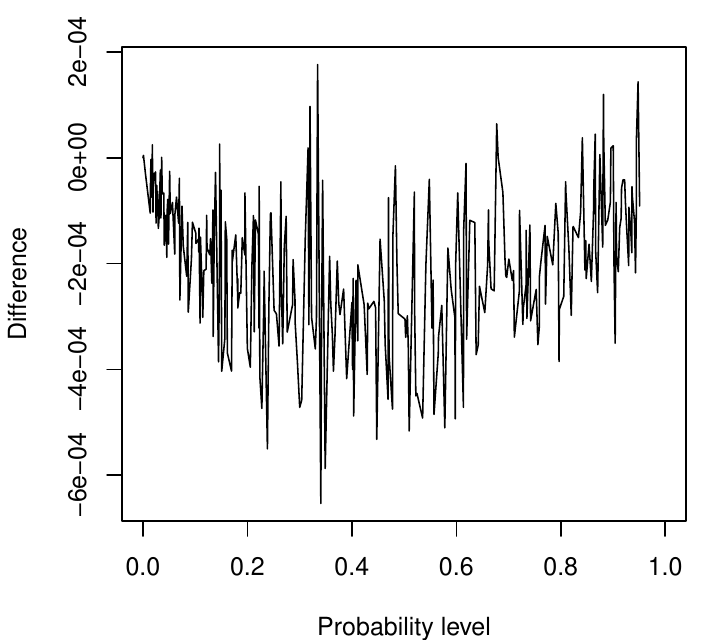}
     \end{subfigure}
    
    \caption{Difference between dense and TLR results of summer wind speed data (on July 15, 2015) in Saudi Arabia.}
    \label{fig:real-data-set2}
\end{figure}

To tackle this issue, we employ Algorithm~\ref{alg:excusion_set} to identify confidence regions. Figure~\ref{fig:real-data-set-c} and Figure~\ref{fig:real-data-set-d} depict these confidence regions, focusing mainly on the mountainous areas in the north, east, and west. Notably, the results obtained from the dense and TLR versions exhibit substantial similarity. We plot the differences between the dense and TLR results across various probability levels to emphasize their distinctions. The analysis reveals that the disparity is of the order of $1 \times 10^{-4}$, further affirming the reliability of the TLR version as shown in Figure~\ref{fig:real-data-set2}.

\subsection{Quantitative Assessment}
Herein, we present the performance evaluation of our Multivariate Normal (MVN) probability implementation on shared- and distributed-memory systems. The comparison includes dense and Tile Low-Rank (TLR) approximations with varying levels of accuracy, considering different MVN dimensions and QMC sample sizes.

\begin{figure}[hbt!]
    \centering
    \begin{subfigure}[b]{0.24\textwidth}
         \centering
         \includegraphics[width=\textwidth,page=1]{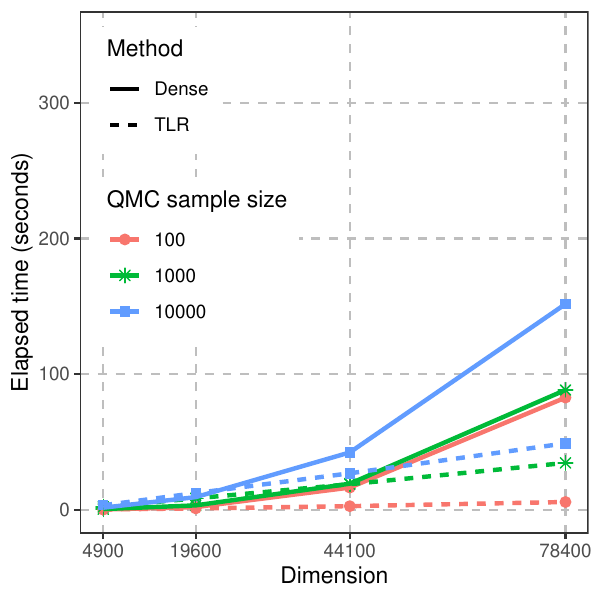}
         \caption{56-core Intel Ice Lake.}
     \end{subfigure}
     \hfill
    \begin{subfigure}[b]{0.24\textwidth}
         \centering
         \includegraphics[width=\textwidth,page=4]{figures/TIME_PMVN_without_arrow.pdf}
         \caption{40-core Intel Cascade Lake.}
     \end{subfigure}
     \hfill
    \begin{subfigure}[b]{0.24\textwidth}
         \centering
         \includegraphics[width=\textwidth,page=3]{figures/TIME_PMVN_without_arrow.pdf}
         \caption{64-core AMD Milan.}
     \end{subfigure}
          \hfill
    \begin{subfigure}[b]{0.24\textwidth}
         \centering
         \includegraphics[width=\textwidth,page=2]{figures/TIME_PMVN_without_arrow.pdf}
         \caption{128-core AMD Naples.}
     \end{subfigure}
    
    \caption{Performance of one MVN integration operation on multiple shared-memory architectures using dense and TLR approximation.}
    \label{fig:perf-shared2}
\end{figure}

\subsubsection{Performance on Shared-Memory Systems}
We leverage four distinct shared-memory architectures to evaluate the time-to-solution of our PMVN algorithm (Algorithm~\ref{alg:pmvn}) {\color{mygreen}as stated in section~\ref{sec:env_set}. We operate on all the cores for each machine}. Performance curves on various architectures, with different Multivariate Normal (MVN) problem dimensions and varying Quasi-Monte Carlo (QMC) sample sizes, are shown in Figure~\ref{fig:perf-shared2}. The dashed curve shows the performance achieved with TLR approximation across various QMC sample sizes, surpassing dense computation by up to $14$X, $19$X, $9$X, and $20$X on Intel Ice Lake, Intel Cascade Lake, AMD Milan, and AMD Naples, respectively, {\color{myred} as shown in table~\ref{tab:perf-shared}. The table also shows that TLR still can achieve better speedup than the dense version with smaller QMC sample sizes, i.e., $100$ and $1000$. }

\begin{table}[hbt!]
    \centering
    \begin{tabular}{|c|c|c|c|}
    \hline
    \multirow{2}{*}{System}       & \multicolumn{3}{|c|}{QMC sample sizes }\\
           \cline{2-4}
     & $100$ & $1000$ & $10{,}000$ \\
    \hline
        56-core Intel Ice Lake & 3X & 3X & 14X \\
        40-core Intel Cascade Lake & 3X & 3X & 19X \\
        64-core AMD Milan & 5X & 5X & 20X \\
        128-core AMD Naples & 2X & 2X & 9X\\
        \hline
    \end{tabular}
    \caption{{\color{myblue}Speedup of TLR to dense implementations with different QMC sample sizes on shared-memory systems.}}
    \label{tab:perf-shared}
\end{table}
The notable speedup achieved through the utilization of TLR compared to dense computation comes from the fast execution of the Cholesky factorization in the TLR format. With a compression accuracy requirement of $1e-3$, validated above through accuracy assessments on synthetic and real datasets, the small ranks of various tiles enable expedited Cholesky factorization compared to the dense version.

\begin{figure*}[t!]
    \centering
    \begin{subfigure}[b]{0.27\textwidth}
         \centering
         \includegraphics[width=\textwidth,page=1]{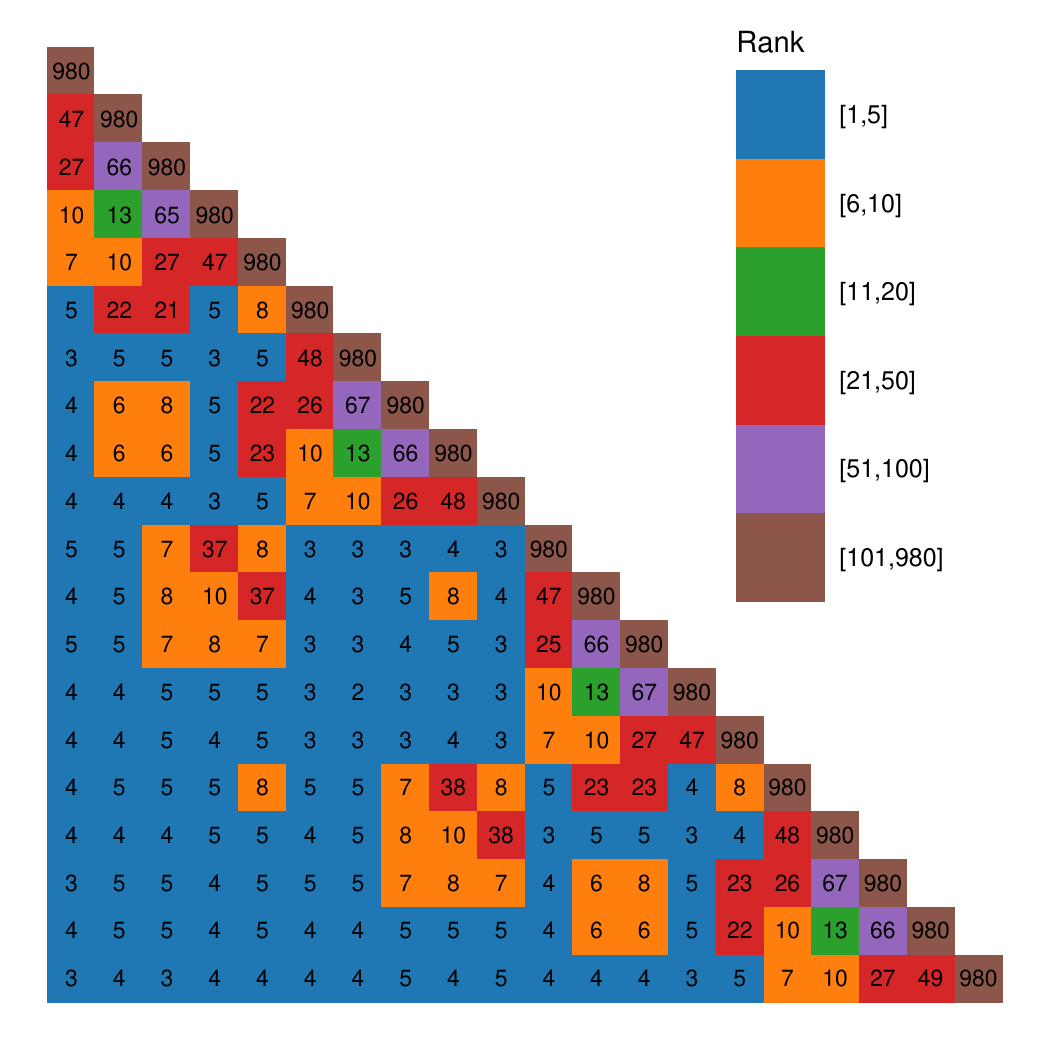}
         \caption{Weak correlation (1, 0.033, 0.5).}
     \end{subfigure}
          \hfill
    \begin{subfigure}[b]{0.27\textwidth}
         \centering
         \includegraphics[width=\textwidth,page=2]{figures/Heatmap_plot.pdf}
         \caption{Medium correlation (1, 0.1, 0.5).}
     \end{subfigure}
     \hfill
    \begin{subfigure}[b]{0.27\textwidth}
         \centering
         \includegraphics[width=\textwidth,page=3]{figures/Heatmap_plot.pdf}
         \caption{Strong correlation (1, 0.234, 0.5).}
     \end{subfigure}
    
    \caption{Rank distributions of a $19600 \times 19600$ covariance matrix using a $980$ tile size with Mat\'ern covariance function under three different settings when compressing the matrix using TLR approximation with accuracy 1e-3 (preserves the application accuracy).}

    \label{fig:ranks}
\end{figure*}
Figure~\ref{fig:ranks} illustrates the rank distributions of a $19{,}600 \times 19{,}600$ matrix at a $1e-3$ compression accuracy. Most tiles exhibit minimal ranks, facilitating a faster computation of the Cholesky factorization operation. The settings employed align with those detailed in Figure~\ref{fig:acc_syn}. Moreover, the figure shows that the ranks exhibit a more pronounced degradation under strong correlations than weak ones, which helps speed up the execution when the correlation is stronger between different locations.

\begin{figure}
         \centering
         \includegraphics[width=0.3\textwidth,page=1]{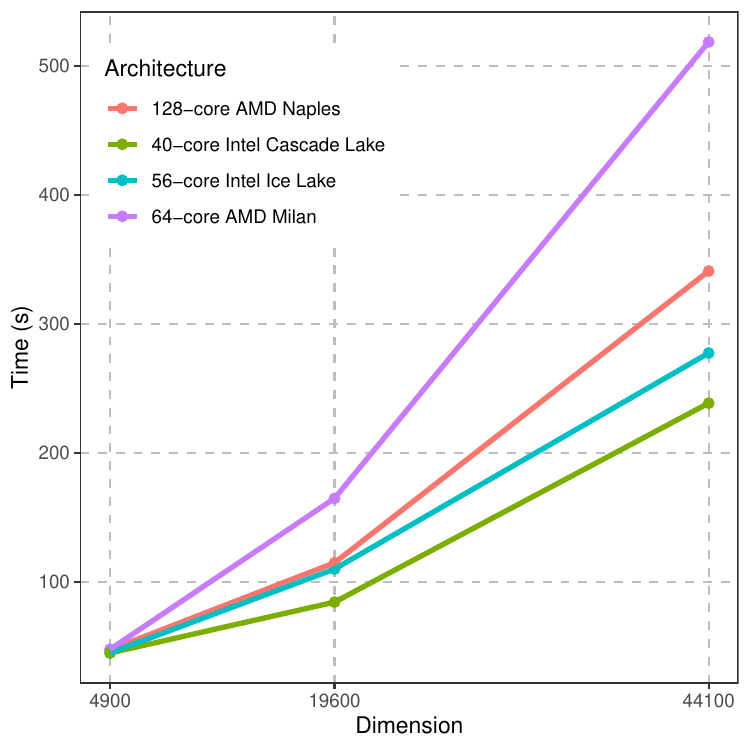}
    \caption{{\color{myblue} Performance of the MC validation process on various dimensional sizes using shared-memory architectures.}}
    \label{fig:MC_validation}
\end{figure}

{\color {myblue} To validate the accuracy of the detected confidence regions, we use an MC validation algorithm as mentioned in~\ref{sec:assess_acc}. The execution overhead of the MC validation process on all four machines above is shown in Figure~\ref{fig:MC_validation} (average over 5 runs). However, because the MC validation is not a complete algorithm to obtain the confidence regions, its execution time should not be considered a part of our algorithm and is unsuitable for comparisons.}

\subsubsection{Performance on Distributed-Memory Systems}
\begin{figure}[h]
    \centering
    \begin{subfigure}[b]{0.24\textwidth}
         \centering
         \includegraphics[width=\textwidth,page=1]{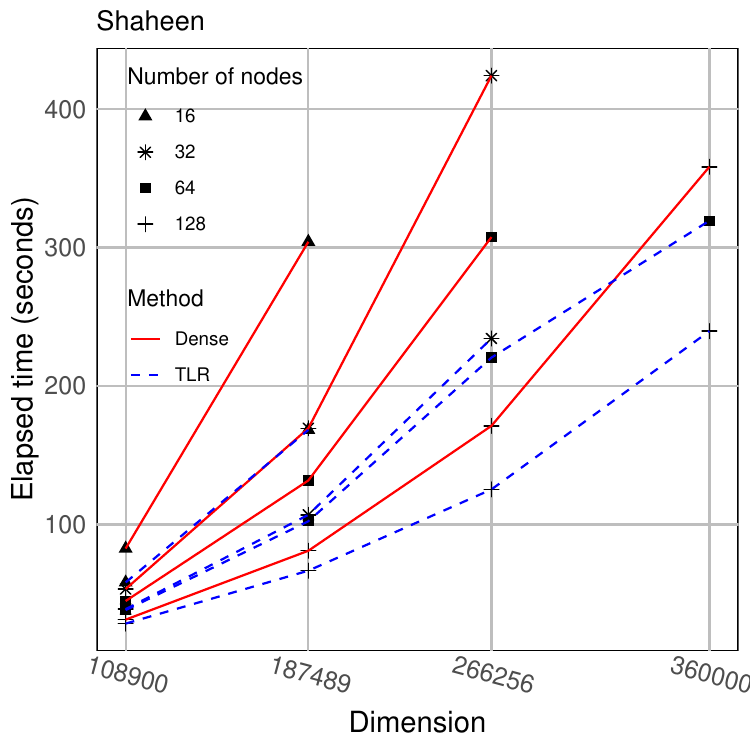}
     \end{subfigure}
          \hfill
    \begin{subfigure}[b]{0.24\textwidth}
         \centering
         \includegraphics[width=\textwidth,page=2]{figures/Shaheen_mvn_perf_without_arrow.pdf}
     \end{subfigure}
    
    \caption{Performance of one MVN integration operation on a Cray XC40 distributed-memory system using dense and TLR approximation.}
    \label{fig:perf-distributed}
\end{figure}
To evaluate the effectiveness of our implementation on distributed-memory systems, we rely on up to 512 nodes of a Cray XC40 system. In Figure~\ref{fig:perf-distributed}, two sub-figures depict the performance on two sets of nodes: 16, 32, 64, and 128, and 64, 128, 256, and 512. We experiment with varying the problem dimensionality across different numbers of nodes. In the left figure, we observe the efficiency of scalability for dense execution (represented by the red curves) across different node configurations, scaling up to $n=360{,}000$. The scalability of TLR (indicated by the dashed blue curves) is also acceptable across various node counts, although some performance degradation is shown when utilizing 64 nodes. {\color{myred}Table~\ref{tab:perf-distributed} shows the speedup of TLR compared to dense using different numbers of nodes.} The speedup of TLR approximation execution compared to the dense computation is up to $1.8$X. In the sub-figure on the right, performance results for up to $512$ nodes and $760{,}384$ data points are presented. The figure demonstrates scalability for both dense and TLR executions, except for the $256$ nodes configuration, which exhibits some performance issues at two points. The maximum speedup achieved by TLR approximation execution compared to the dense execution in this sub-figure is $1.7$X.
\begin{table}[]
    \centering
    \begin{tabular}{|c|c|}
    \hline
    \multirow{2}{*}{Number of nodes} & Speedup using QMC \\
    & sample size = $10{,}000$ \\
    \hline
        16 &  1.8X\\
        32 & 1.8X \\
        64 & 1.4X \\
        128 & 1.7X \\
        256 & 1.3X \\
        512 & 1.5X \\
        \hline
    \end{tabular}
    \caption{{\color{myblue}Speedup of TLR to dense implementations on a Cray XC40 system with different number of nodes. The most accurate QMC sample size is used, i.e., $10{,}000$.}}
    \label{tab:perf-distributed}
\end{table}

Comparing the performance with shared-memory computations, the dense computation in the MVN probability Algorithm~\ref{alg:pmvn} takes more time than the Cholesky factorization operation, which is the only part that can be computed in either dense or low-rank format. This is why the differential impact on performance is observed in the TLR version of the algorithm on distributed-memory systems. According to our experiments, the TLR Cholesky factorization within the MVN probability algorithm only can achieve speedups of $5.2$X, $4.5$X, $2.6$X, $3.1$X, $1.9$X, and $2.6$X on $16$, $32$, $64$, $128$, $256$, and $512$ nodes, respectively, when compared to the dense version.


\section{Related Work}
Multivariate normal (MVN) probability frequently appears in statistical applications, including the probability density functions of several skew-normal~\cite{gonzalez2019shapiro}, Bayesian probit models~\cite{anceschi2023bayesian}, and confidence regions detection problems~\cite{bolin2015excursion}, and maximum likelihood estimation~\cite{cao2021exploiting}. It can be defined as a numerical integration problem with $n$ dimensions, which has a prohibitive computation when $n$ is large. Computing the MVN probability is a challenging task where quadrature-based algorithms are impractical in higher dimensions.
In 1992, Genz et al. \cite{genz1992numerical} proposed a Monte Carlo-based solution to reduce the complexity of computing the MVN probability by applying a set of transformations to the original problem. The proposed algorithm has $O(n^3)$ complexity to compute the Cholesky factorization of $n \times n$ covariance matrix and $O(n^2)$ to process a single MC sample. In the literature, many solutions have been provided. For instance, Genton et al. \cite{genton2018hierarchical} have proposed a hierarchical quasi-Monte Carlo (QMC) method to improve the underlying linear algebra operations to compute the MVN probability. The provided optimization includes compressing the covariance matrix using a hierarchical low-rank approximation to reduce the complexity of processing a single MC sample from $O(n^2)$ to $O(mn+kn\log(n/m))$, i.e., $k$ is the rank of off-diagonal matrix blocks and $m$ is the inadmissible matrix blocks. In~\cite{botev2017normal}, the minimax tilting method has been proposed as an efficient approximation to the MVN problem, accurately estimating the required probability. The minimax tilting method can improve the convergence rate but requires an optimization step to $O(n)$ parameters. In~\cite{azzimonti2018estimating}, a two-step method has been proposed where the original MVN probability is decomposed into a low-dimensional and high-dimensional residual. However, this method can be used in the case of orthant MVN problems with constant upper and lower integration limits. 
The Multivariate Normal (MVN) probability finds utility in various applications, including detecting confidence regions. In such applications, the key objective is identifying regions surpassing a specified threshold within a given confidence level. These applications are prevalent in medical~\cite{ejaz2020hybrid,ejaz2022confidence} and environmental sciences~\cite{barnett2001detection,marsili2003confidence}.

\section{Conclusion}

This study presents a novel mitigation of the problem of the high complexity of utilizing MVN by proposing a parallel implementation of the SOV algorithm designed for manycore systems. This implementation enables rapid computation of MVN probability using dense and Tile Low-Rank (TLR) approximation methods by relying on the StarPU dynamic runtime system and advanced linear algebra libraries, i.e., Chameleon and HiCMA. Our proposed implementation has been applied to the context of confidence region detection applications, demonstrating its effectiveness in handling large spatial regions encompassing up to $700$K locations. Furthermore, our proposed TLR-based implementations exhibit a notable speedup compared to the dense solution, achieving up to 20X improvement on shared-memory systems. These enhancements come with high accuracy achievements compared to dense solutions. We assess the accuracy of our implementation through confidence region detection applications using synthetic and wind speed datasets.

In future work, we aim to incorporate multi- and mixed-precision executions to enhance support for the MVN probability algorithm. The results obtained in this paper indicate that the algorithm maintains the necessary accuracy even at very low levels of TLR compression accuracy. Consequently, we anticipate that lower-precision executions can expedite the computation process with minimal impact on the required accuracy. In such a scenario, the natural extension of this work would involve GPU support, leveraging tensor core execution for the underlying linear algebra operations.

\section{Acknowledgment}
Financial support and backing for this study were provided by King Abdullah University of Science and Technology (KAUST) via the Office of Sponsored Research (OSR). The research utilized the facilities of the Extreme Computing Research Center (ECRC) and the KAUST Supercomputing Laboratory (KSL). Key resources employed in this study included the Cray XC40 and the Shaheen II supercomputer, which are pivotal assets of the KSL.
\bibliographystyle{IEEEtran}
\bibliography{ref}

\end{document}